Specific-Heat Study of Superconducting and Normal States in FeSe$_{1-x}$Te$_x$ (0.6 ≤ x ≤ 1) Single Crystals: Strong-Coupling Superconductivity, Strong Electron-Correlation, and Inhomogeneity


Takashi Noji, Masato Imaizumi, Takumi Suzuki[*], Tadashi Adachi, Masatsune Kato, and Yoji Koike

Department of Applied Physics, Tohoku University, Sendai 980-8579, Japan



The electronic specific heat of as-grown and annealed single-crystals of FeSe$_{1-x}$Te$_x$ (0.6 ≤ x ≤ 1) has been investigated. It has been found that annealed single-crystals with x = 0.6 - 0.9 exhibit bulk superconductivity with a clear specific-heat jump at the superconducting (SC) transition temperature, $T_c$. Both $2\Delta_0/k_BT_c$ [$\Delta_0$: the SC gap at 0 K estimated using the single-band BCS s-wave model] and $\triangle C/(\gamma_n-\gamma_0)T_c$ [$\triangle C$: the specific-heat jump at $T_c$, $\gamma_n$: the electronic specific-heat coefficient in the normal state, $\gamma_0$: the residual electronic specific-heat coefficient at 0 K in the SC state] are largest in the well-annealed single-crystal with x = 0.7, i.e., 4.29 and 2.76, respectively, indicating that the superconductivity is of the strong coupling. The thermodynamic critical field has also been estimated. $\gamma_n$ has been found to be one order of magnitude larger than those estimated from the band calculations and increases with increasing x at x = 0.6 - 0.9, which is surmised to be due to the increase in the electronic




effective mass, namely, the enhancement of the electron correlation. It has been found that there remains a finite value of $\gamma_0$ in the SC state even in the well-annealed single-crystals with x = 0.8 - 0.9, suggesting an inhomogeneous electronic state in real space and/or momentum space.



## 1. Introduction

The study of iron-based superconductors triggered by the discovery of superconductivity in the iron-pnictide $LaFeAsO_{1-x}F_x$ has been expanding and has remained active,[1] because the superconducting (SC) transition temperature, $T_c$, of $SmFeAsO_{1-x}F_x$ is as high as 55 K[2] and there is a wide variety of crystal structures in iron-based superconductors as in the case of copper-based superconductors. Among iron-based superconductors, $FeSe_{1-x}Te_x$ has attracted great interest,[3] because it has the simplest crystal structure composed of a stack of edge-sharing $Fe(Se,Te)_4$-tetrahedra layers which are similar to $FeAs_4$-tetrahedra layers in $LaFeAsO_{1-x}F_x$. According to band calculations,[4,5] moreover, the Fermi surface of $FeSe_{1-x}Te_x$ is similar to that of iron pnictides. In fact, an angle-resolved photoemission spectroscopy (ARPES) study has revealed that both quasi-two-dimensional hole and electron pockets exist at the



Brillouin-zone center and edge, respectively, which are similar to those of iron pnictides.[6,7)] Therefore, if the Fermi surface topology is related to the mechanism of the appearance of superconductivity, the mechanism in $FeSe_{1-x}Te_x$ might be the same as that in iron pnictides.

The $T_c$ of $FeSe_{1-x}Te_x$ has been found to increase from 8 K in FeSe with increasing x and to show a maximum 14 K at x ~ 0.6 in $FeSe_{1-x}Te_x$, and the superconductivity of $FeSe_{1-x}Te_x$ disappears at x = 1, namely, in FeTe.[8,9)] The end member FeTe is not SC but it develops an antiferromagnetic (AF) order at low temperatures below ~ 67 K, at which a tetragonal-to-monoclinic structural phase transition occurs.[10,11)] As for single crystals of $FeSe_{1-x}Te_x$ ($0.5 \leq x \leq 1$), Sales *et al.*[12)] have reported that only single crystals with x ~ 0.5 exhibit bulk superconductivity from electrical-resistivity, magnetic-susceptibility, and specific-heat measurements. On the other hand, our magnetic-susceptibility measurements have revealed that single crystals with a wide range of x = 0.5 - 0.9 exhibit bulk superconductivity through annealing at 400℃ for 100 h in vacuum.[13)] Moreover, it has been found that the magnetic transition observed in as-grown single-crystals with a range of x = 0.8 - 0.9 disappears through the annealing and that the magnetic transition takes place only around x = 1 in the annealed single-crystals.[13)] According to the EDX analysis in $FeSe_{0.39}Te_{0.61}$ by Taen et al.,[14)] the distribution of Se and Te in the crystal becomes homogeneous through the annealing. The width of the powder x-ray diffraction peaks of our single-crystals with x = 0.6 - 1



has decreased through the annealing.[15] Moreover, the in-plane electrical resistivity at low temperatures has been found to become more metallic through the annealing in our single-crystals with x = 0.5 - 1. Therefore, it appears that the annealing makes the distribution of Se and Te in a crystal homogeneous and relaxes lattice distortion, leading to the appearance of bulk superconductivity in a wide range of x = 0.5 - 0.9.

In this paper, we have investigated the electronic specific heat of as-grown and annealed single-crystals of FeSe$_{1-x}$Te$_x$ ($0.6 \leq x \leq 1$), to clarify the nature of the bulk SC state as well as the electronic state in the normal state. For the annealed single-crystals exhibiting bulk superconductivity, the SC condensation energy at 0 K, U$_0$, and the SC gap at 0 K, $\Delta_0$, have been estimated. The thermodynamic critical field, H$_c$(T), has also been estimated. The variations of the electronic specific-heat coefficient in the normal state, $\gamma_n$, and of the residual electronic specific-heat coefficient at 0 K in the SC state, $\gamma_0$, with x have been observed and discussed in terms of the electronic effective mass and inhomogeneity in real space and/or momentum space, respectively.

2. Experimental

Single crystals of FeSe$_{1-x}$Te$_x$ were grown by the Bridgman method. Powders of Fe, Se, and Te, prescribed in the nominal composition, were thoroughly mixed in an argon-filled glove box and sealed in an evacuated quartz tube. Since the quartz tube often cracked upon cooling, the tube was sealed in another large-sized evacuated quartz



tube. The doubly sealed quartz ampoule was placed in a furnace, heated up to 950 -

1050℃, and cooled. As-grown crystals obtained thus were annealed at 400℃ for 100 -

200 h in vacuum. Crystals were characterized by x-ray back-Laue photography and

powder x-ray diffraction analysis. The chemical composition was determined by

inductively coupled plasma optical emission spectroscopy (ICP-OES). The composition

of the surface of the crystals was checked using an electron probe microanalyzer

(EPMA). The details are described in our previous paper.[13] The specific heat was

measured by the thermal-relaxation method at low temperatures down to 2 K, using a

commercial apparatus (Quantum Design, PPMS).

3. Results and Discussion

Figure 1 shows the temperature dependence of the specific heat, C, of

$FeSe_{1-x}Te_x$ ($0.6 \leq x \leq 1$) single crystals as-grown and annealed at 400℃ for 100 and 200

h, plotted as C/T vs $T^2$. It is found that a jump in the specific heat at $T_c$ is clearly

observed in the 100h- and 200h-annealed crystals with x = 0.6 - 0.9, indicating that bulk

superconductivity appears in these crystals. Among the as-grown crystals, a slight

specific-heat jump at $T_c$ is observed at $T^2 \sim 50$ $K^2$ only in the crystal with x = 0.6. These

are consistent with the results of our magnetic-susceptibility measerements.[13] As far as

the specific heat in the normal state is concerned, it appears that there is no large

difference between the as-grown and annealed crystals.



To estimate the electronic specific heat, $C_{el}$, the phonon specific heat, $C_{ph}$, must be subtracted from the total specific heat shown in Fig. 1. To obtain an accurate $C_{ph}$, therefore, we prepared a non-SC 100h-annealed single-crystal of $Fe_{0.95}Cu_{0.05}Se_{0.4}Te_{0.6}$ in which the superconductivity is suppressed through the partial substitution of Cu for Fe, using the same technique of growth and annealing as the preparation of the 100h-annealed single-crystals of $FeSe_{1-x}Te_x$. The temperature dependence of the specific heat, C, of the 100h-annealed single-crystal of $Fe_{0.95}Cu_{0.05}Se_{0.4}Te_{0.6}$, plotted as $C/T$ vs $T^2$, is also shown in Fig. 1. No specific-heat jump is observed owing to the complete suppression of superconductivity, so that the specific heat of $Fe_{0.95}Cu_{0.05}Se_{0.4}Te_{0.6}$ is suitable for estimating $C_{ph}$ in $FeSe_{1-x}Te_x$. The specific heat of $Fe_{0.95}Cu_{0.05}Se_{0.4}Te_{0.6}$ was well fitted using

$$C = \gamma_n T + \beta T^3 + \delta T^5 + \varepsilon T^7, \qquad (1)$$

where $\gamma_n T$ is $C_{el}$ and $\beta T^3 + \delta T^5 + \varepsilon T^7$ is $C_{ph}$. The values of the fitting parameters are listed in Table I. Next, the $C_{ph}$ of the SC annealed single-crystals of $FeSe_{1-x}Te_x$ ($0.6 \leq x \leq 0.9$) was estimated as $\beta(\alpha T)^3 + \delta(\alpha T)^5 + \varepsilon(\alpha T)^7$, using the values of $\beta$, $\delta$, and $\varepsilon$ obtained by the above fitting of C in $Fe_{0.95}Cu_{0.05}Se_{0.4}Te_{0.6}$. Here, $\alpha$ is a parameter reflecting the difference in atomic mass between $Fe_{0.95}Cu_{0.05}Se_{0.4}Te_{0.6}$ and $FeSe_{1-x}Te_x$ and is nearly unity. Then, the $C_{el}$ of the SC annealed single-crystals was obtained by subtracting the $C_{ph}$ estimated from the total specific heat, C, as

$$C_{el} = C - \{\beta(\alpha T)^3 + \delta(\alpha T)^5 + \varepsilon(\alpha T)^7\}. \qquad (2)$$



$\gamma_n$ was also estimated as $C_{el}/T$ above $T_c$. Here, $\alpha$ and $\gamma_n$ were determined, taking into account the so-called entropy balance by which the electronic entropy in the SC state, $S^S$, accords with that in the normal state, $S^n$, at the onset temperature of superconductivity, $T_c^{onset}$, namely,

$$\int_0^{T_c^{onset}} \frac{C_{el}}{T}\, dT = \gamma_n T_c^{onset} \,. \qquad (3)$$

The temperature dependence of the thus-obtained $C_{el}$ divided by temperature for the 200h-annealed single-crystals of $FeSe_{1-x}Te_x$ ($0.6 \leq x \leq 0.9$) is shown in Fig. 2. Excess specific-heat, which is due to the large SC fluctuation caused by the low-dimensionality of superconductivity, is observed just above $T_c$ for all the crystals. $T_c$ was determined as shown in the figure, taking into account the SC fluctuation. The excess specific-heat above ~15 K at $x = 0.9$ may be due to magnetism related to the AF ordering around $x = 1$. Both $\gamma_n$ and $\gamma_0$, defined as the values of $C_{el}/T$ extrapolated to 0 K, are also indicated in Fig. 2. The values of $\gamma_n$, $\gamma_0$, $\beta$, $\delta$, $\varepsilon$, and $\alpha$ of annealed single-crystals of $FeSe_{1-x}Te_x$ thus determined are also listed in Table I. As for 100h- and 200h-annealed non-SC crystals with $x = 0.95$ and 1, these values were determined using eq. (1), as in the case of $Fe_{0.95}Cu_{0.05}Se_{0.4}Te_{0.6}$.

Figure 3 shows the x dependences of $\gamma_n$ and $\gamma_0$ for $FeSe_{1-x}Te_x$ ($0.6 \leq x \leq 1$) single-crystals annealed at 400℃ for 100 and 200 h. $\gamma_n$ corresponds to the electronic density of states at the Fermi level in the normal state, while $\gamma_0$ corresponds to the electronic density of states at the Fermi level still remaining at 0 K in the SC state.



It is found that $\gamma_n$ increases with increasing x and suddenly decreases above x = 0.9. The decrease above 0.9 is surmised to be caused by the partial disappearance of the Fermi surface due to the AF ordering and/or the structural phase transition around x = 1. As to the increase in $\gamma_n$ with increasing x in the range of x = 0.6 - 0.9, it is expected to be due to the increase in the carrier density or electronic effective mass. However, it does not appear that the carrier density increases with increasing x, because the in-plane electrical resistivity in the normal state tends to increase with increasing x. The $\gamma_n$ values are in good agreement with those at x = 0.6 and 1 reported so far [12,16-19] and comparable with those of other almost optimally doped iron-based superconductors, namely, 39 mJ/molK$^2$ in SmFeAsO$_{0.85}$F$_{0.15}$,[20] 23.8 mJ/molK$^2$ in Ba(Fe$_{0.925}$Co$_{0.075}$)$_2$As$_2$,[21] and 53 mJ/molK$^2$ in Ba$_{0.7}$K$_{0.3}$Fe$_2$As$_2$,[22] but they are one order of magnitude larger than those estimated from the band calculations, namely, 3.055 mJ/molK$^2$ in FeSe and 4.783 mJ/molK$^2$ in FeTe.[5] This suggests that the electronic effective mass shows a larger increase than the band mass. In fact, the increase in the effective mass, $m^*/m_{band}$, has been estimated from the ARPES study to be 6 - 20 in FeSe$_{0.42}$Te$_{0.58}$ by Tamai et al.,[23] which is consistent with our specific-heat results. Accordingly, it is likely that the effective mass in FeSe$_{1-x}$Te$_x$ is large and increases further toward a possible quantum critical point with increasing x, leading to an increase in $\gamma_n$ with increasing x at x = 0.6 - 0.9 and the appearance of the AF order above x = 0.9. This is reasonable, because the chalcogen height from the iron plane in the



Fe(Se,Te)$_4$-tetrahedra layer increases with increasing x so that the electronic band width becomes narrow and the electron correlation expectedly becomes enhanced with increasing x.[24] Therefore, it is possible that the spin fluctuations and/or orbital fluctuations are large and are further enhanced with increasing x at x = 0.6 - 0.9.[25-28] In fact, strong spin fluctuations have been observed in neutron scattering[29,30] and NMR[31] measurements of FeSe$_{1-x}$Te$_x$. Note, however, that the direction of the in-plane wave vector of the static AF order at x = 1 is 45° away from that of the spin fluctuations.[10,11]

As for $\gamma_0$, it is found to be nearly zero in the 200h-annealed single-crystals with x = 0.6 and 0.7, indicating that there is no normal-state region in these crystals at 0 K. It is speculated that the distribution of Se and Te in a crystal has become homogeneous and the lattice distortion has been relaxed through the long annealing so that the SC state has become homogeneous in the 200h-annealed crystals with x = 0.6 - 0.7. Finite values of $\gamma_0$ observed around x = 0.6 by some groups[18,19] will be due to insufficiency of annealing. For x = 0.8 and 0.9, on the other hand, there remain finite values of $\gamma_0$ even in the 200h-annealed crystals. Moreover, the $\gamma_0$ values of the 200h-annealed crystals are not so different from those of the 100h-annealed crystals. Therefore, the finite values of $\gamma_0$ seem to be intrinsic at x = 0.8 - 0.9, indicating that normal-state electrons exist at 0 K even in annealed SC crystals of x = 0.8 - 0.9. The finite values of $\gamma_0$ at x = 0.8 - 0.9 may be interpreted as being due to inhomogeneity in real space, namely, due to a phase separation into SC and normal-state regions. On the other hand, they may be due to an



inhomogeneity in momentum space, because the Fermi surface of these compounds is composed of multi-pockets of electrons and holes. That is, part of the Fermi surface may remain in the normal state even at 0 K at x = 0.8 - 0.9. Otherwise, the electronic state may be inhomogeneous in both real and momentum spaces. To be conclusive, further investigation is necessary.

Next, we estimate several SC parameters of the annealed single-crystals exhibiting bulk superconductivity from the data of $C_{el}$ shown in Fig. 2. The SC condensation energy at 0 K, $U_0$, can be estimated using

$$U_0 = \int_0^{T_c^{onset}} \left( S^n - S^s \right) dT = \frac{\gamma_n}{2} \left( T_c^{onset} \right)^2 - \int_0^{T_c^{onset}} \left( \int_0^T \frac{C_{el}}{T} \, dT \right) dT \, . \qquad (4)$$

For example, the temperature dependences of $S^n$ and $S^s$ of the 200h-annealed crystal with x = 0.7 is obtained from the data of $C_{el}/T$ in Fig. 2, as shown in Fig. 4. The value of $U_0$ is given by the area surrounded by $S^n$ and $S^s$ in the figure. The x dependence of $U_0$ thus obtained is shown in Fig. 5(b). The values of $U_0$ are 1.74 and 2.53 J/mol in the 200h-annealed single-crystals with x = 0.6 and 0.7, respectively, and tend to decrease in those with x = 0.8 - 0.9.

The thermodynamic critical field at 0K, $H_c(0)$, is estimated using

$$U_0 = \frac{V}{8\pi} H_c^2(0) \, , \qquad (5)$$

where V is the volume. The x dependence of $H_c(0)$ is shown in Fig. 5(c). The values of $H_c(0)$ are 0.406 and 0.486 T in the 200h-annealed single crystals with x = 0.6 and 0.7, respectively. As for the single crystals whose $\gamma_0$ values are finite, $H_c(0)$ may be



underestimated, because V in eq. (5) may actually be reduced due to the possible inhomogeneous superconductivity.

The SC gap at 0 K, $\Delta_0$, is estimated using the following equation based on the single-band BCS s-wave model,

$$\Delta_0 = \left\{ \frac{4\pi^2 k_B^2 U_0}{3(\gamma_n - \gamma_0)} \right\}^{1/2} , \qquad (6)$$

because

$$U_0 = \frac{1}{2} N(0) \Delta_0^2 , \qquad (7)$$

$$N(0) = \frac{3(\gamma_n - \gamma_0)}{2\pi^2 k_B^2} . \qquad (8)$$

Here, N(0) is the density of states at the Fermi level of electrons in the normal state contributing to the superconductivity and $k_B$ is the Boltzmann constant. The x dependence of $\Delta_0$ thus obtained is shown in Fig. 5(d). The values of $\Delta_0$ are 2.54 and 2.63 meV in the 200h-annealed single-crystals with x = 0.6 and 0.7, respectively, which are comparable with those obtained from the muon spin relaxation of FeSe$_{0.5}$Te$_{0.5}$ by Biswas et al.,[32] and Bendele et al.,[33] from the specific heat of FeSe$_{0.5}$Te$_{0.5}$ by Tsurkan et al.,[34] and from the optical conductivity of FeSe$_{0.45}$Te$_{0.55}$ by Homes et al.[35] $\Delta_0$ as well as U$_0$ tends to decrease at x = 0.8 - 0.9. $2\Delta_0/k_BT_c$ is estimated as 4.24 and 4.29 in the 200h-annealed crystals with x = 0.6 and 0.7, respectively, as shown in Fig. 5(e). Here, T$_c$ is defined as shown in Fig. 2. These values of $2\Delta_0/k_BT_c$ are in rough agreement with



those obtained from the specific heat of $FeSe_{0.43}Te_{0.57}$ by Hu *et al.*[19] and much larger than 3.52 owing to the BCS weak-coupling theory, indicating that the superconductivity is of the strong coupling. Note that the $2\Delta_0/k_BT_c$ values smaller than 3.52 obtained from the specific-heat measurements by some groups[12,18] may be due to the ambiguity in the estimate of $C_{ph}$ in SC crystals of $FeSe_{1-x}Te_x$. In almost optimally doped samples of both Fe122 and Fe1111 systems, large values of $2\Delta_0/k_BT_c$ suggesting strong-coupling superconductivity have also been obtained from the specific-heat measurements.[20-22,36,37]

From the specific heat jump at $T_c$, $\triangle C$, the maximum and minimum values of $\triangle C/T_c$ are defined as shown in Fig. 2. Then, $\triangle C/(\gamma_n-\gamma_0)T_c$ is estimated as shown in Fig. 5(f). The values of $\triangle C/(\gamma_n-\gamma_0)T_c$ are 2.11 and 2.76 in the 200h-annealed single-crystals with x = 0.6 and 0.7, respectively. These values are in agreement with those obtained by Hu *et al.*[19] and are much larger than 1.43 owing to the BCS weak-coupling theory, also indicating strong-coupling superconductivity. $\triangle C/(\gamma_n-\gamma_0)T_c$ tends to decrease as well as $U_0$ and $\Delta_0$ at x = 0.8 − 0.9. As for the $\triangle C/(\gamma_n-\gamma_0)T_c$ values of almost optimally doped samples of Fe122 and Fe1111 systems, those of the Fe122 system are slightly larger 1.43,[38,39] while those of the Fe1111 system are much smaller than 1.43.[20,40] The latter may be due to the poor quality of the polycrystalline samples.

To summarize the results shown in Fig. 5, the superconductivity is strongest in the 200h-annealed single-crystal with x = 0.7, where it is of the strong coupling. At x =



0.8 - 0.9, the superconductivity becomes weaker possibly owing to normal-state electrons still remaining in the SC state.

Finally, we estimate the temperature dependence of $H_c(T)$ from the SC condensation energy at each temperature, $U(T)$. $U(T)$ can be estimated from the data of $C_{el}$ shown in Fig. 2, using

$$U(T) = \int_T^{T_c^{onset}} \left( S^n - S^s \right) dT \, . \qquad (9)$$

$U(0)$ is the same as $U_0$ in eq. (4). Then using

$$U(T) = \frac{V}{8\pi} H_c^2(T) \, , \qquad (10)$$

$H_c(T)$ is estimated as shown in Fig. 6. As for the 200-annealed crystals with x = 0.8 and 0.9, $H_c(T)$ may be underestimated, because V in eq. (10) may actually be reduced due to the possible inhomogeneous superconductivity. The dashed line in Fig. 6 indicates the empirical parabolic law,

$$H_c(T) = H_c(0) \left\{ 1 - \left( \frac{T}{T_c} \right)^2 \right\} \, , \qquad (11)$$

putting $H_c(0)$ at the estimated value at 0K. It is found that the estimated values of $H_c(T)$ deviate from the parabolic law at high temperatures above ~ 0.5$T_c$. The deviation around and above $T_c$ is due to the large SC fluctuation, as mentioned above. The deviation far below $T_c$ may be due to the strong-coupling superconductivity[41] and/or multi-band superconductivity.



## 4. Summary


We have investigated the electronic specific heat of $FeSe_{1-x}Te_x$ ($0.6 \leq x \leq 1$) single crystals as-grown and annealed at 400℃ for 100 - 200 h in vacuum. The appearance of bulk superconductivity with a clear specific-heat jump at $T_c$ has been confirmed in the annealed single-crystals with x = 0.6 - 0.9. For the annealed single-crystals with x = 0.6 - 0.9, $U_0$, $H_c(0)$, $\Delta_0$ and $\triangle C/(\gamma_n-\gamma_0)T_c$ have been estimated. It has been found that the superconductivity is strongest in the 200h-annealed single-crystal with x = 0.7, where $2\Delta_0/k_BT_c$ estimated using the single-band BCS s-wave model and $\triangle C/(\gamma_n-\gamma_0)T_c$ are 4.29 and 2.76, respectively, indicating that the superconductivity is of the strong coupling. The temperature dependence of $H_c(T)$ has been estimated and found to be larger than that determined by the empirical parabolic law, which may be due to the strong-coupling superconductivity and/or multi-band superconductivity. $\gamma_n$ has been estimated to be one order of magnitude larger than those estimated from the band calculations and to increase with increasing x at x = 0.6 - 0.9, suggesting that the electronic effective mass is much larger than the band mass and increases with increasing x. That is, it appears that the electron correlation inducing spin fluctuations and/or orbital fluctuations is not weak and is enhanced with increasing x at x = 0.6 - 0.9. $\gamma_n$ decreases around x = 1 owing to the AF ordering and/or the structural phase transition. $\gamma_0$ is nearly zero in the 200h-annealed single-crystals with x = 0.6 - 0.7, indicating that the SC state is homogeneous at x = 0.6 - 0.7. For x = 0.8 - 0.9, on the




other hand, there remain finite values of $\gamma_0$ even in the 200h-annealed crystals, indicating that normal-state electrons remain even at 0 K in the SC state. This result suggests an inhomogeneous electronic state in real space, namely, a phase separation into the SC and normal-state regions, and/or that in momentum space, namely, the coexistence of SC and normal-state pockets at the Fermi level even in the SC state.


Acknowledgments

We would like to thank N. L. Saini, T. Mizokawa, T. Tohyama, K. Kuroki and T. Kawamata for helpful discussion. We are grateful to K. Takada and M. Ishikuro of Institute for Materials Research (IMR), Tohoku University, for their aid in the ICP-OES analysis. We are also indebted to Y. Murakami of the Advanced Research Center of Metallic Glasses, IMR, Tohoku University, for his aid in the EPMA measurements. This work was supported by a Grant-in-Aid for Scientific Research from the Japan Society for the Promotion of Science.




References


[*]Present address: Institute for Materials Research, Tohoku University, Sendai 980-8577

083914.

Figure captions

Fig. 1. (Color online) Temperature dependence of the specific heat, C, of $FeSe_{1-x}Te_x$ ($0.6 \leq x \leq 1$) single crystals as-grown and annealed at 400℃ for 100 and 200 h and of a $Fe_{0.95}Cu_{0.05}Se_{0.4}Te_{0.6}$ single crystal annealed at 400℃ for 100 h, plotted as C/T vs $T^2$.

Fig. 2. (Color online) Temperature dependence of the electronic specific heat divided by temperature, $C_{el}/T$, for $FeSe_{1-x}Te_x$ ($0.6 \leq x \leq 1$) single crystals annealed at 400℃ for 200 h. Both the electronic specific-heat coefficient in the normal state, $\gamma_n$, and the residual electronic specific-heat coefficient at 0 K in the SC state, $\gamma_0$, defined as the $C_{el}/T$ extrapolated to 0 K are shown. Definitions of $T_c$ and the specific-heat jump at $T_c$ divided by $T_c$, $\triangle C/T_c$, are also shown.

Fig. 3. (Color online) Dependences on x of the electronic specific-heat coefficient in the normal state, $\gamma_n$, and the residual electronic specific-heat coefficient at 0 K in the SC state, $\gamma_0$, for $FeSe_{1-x}Te_x$ ($0.6 \leq x \leq 1$) single crystals annealed at 400℃ for 100 and 200 h.

Fig. 4. (Color online)Temperature dependences of the electronic entropy in the normal state, $S^n$, and that in the SC state, $S^s$.



Fig. 5. (Color online)　Dependences on x of (a) the SC transition temperature, $T_c$, (b) the SC condensation energy at 0 K, $U_0$, (c) the thermodynamic critical field at 0K, $H_c(0)$, (d) the SC gap at 0 K, $\Delta_0$, estimated using the single-band BCS s-wave model, (e) $2\Delta_0/k_BT_c$, and (f) $\triangle C/(\gamma_n-\gamma_0) T_c$ for FeSe$_{1-x}$Te$_x$ ($0.6 \le x \le 1$) single crystals annealed at 400℃ for 100 and 200 h. Here, $T_c$ is defined as shown in Fig. 2. $\triangle C$ is the specific-heat jump at $T_c$, and $\triangle C/T_c$ is defined as shown in Fig. 2. Single crystals of FeSe$_{1-x}$Te$_x$ with $x \geqq 0.95$ show no specific-heat jump above 2 K.

Fig. 6. (Color online) Temperature dependence of the thermodynamic critical field, $H_c(T)$, for FeSe$_{1-x}$Te$_x$ ($0.6 \le x \le 0.9$) single crystals annealed at 400℃ for 200 h. Dashed lines indicate the empirical parabolic law of $H_c(T) = H_c(0)\{ 1-(T/T_c)^2\}$.



Table I. Parameters used for the fitting of the temperature dependence of the specific heat for the 100h-annealed $Fe_{0.95}Cu_{0.05}Se_{0.4}Te_{0.6}$ single crystal and 100h- and 200h-annealed $FeSe_{1-x}Te_x$ single crystals. The values of $\beta$, $\delta$, and $\varepsilon$ of the 100h- and 200h-annealed $FeSe_{1-x}Te_x$ with x = 0.6 - 0.9 were fixed to be those of the100h-annealed $Fe_{0.95}Cu_{0.05}Se_{0.4}Te_{0.6}$.

| | x | $\gamma_n$ (mJ/molK$^2$) | $\gamma_0$ (mJ/molK$^2$) | $\beta$ (mJ/molK$^4$) | $\delta$ (mJ/molK$^6$) | $\varepsilon$ (mJ/molK$^8$) | $\alpha$ | $T_c$ (K) |
|---|---|---|---|---|---|---|---|---|
| $Fe_{0.95}Cu_{0.05}Se_{0.4}Te_{0.6}$ (100h-annealed) | 0.6 | 22.9 | 22.9 | 0.782 | $-8.74\times10^{-4}$ | $3.86\times10^{-7}$ | — | — |
| $FeSe_{1-x}Te_x$ (100h-annealed) | 0.6 | 28.0 | 0.4 | 0.782 | $-8.74\times10^{-4}$ | $3.86\times10^{-7}$ | 0.943 | 14.0 |
| | 0.7 | 48.2 | 19.8 | 0.782 | $-8.74\times10^{-4}$ | $3.86\times10^{-7}$ | 0.981 | 14.6 |
| | 0.8 | 60.1 | 28.0 | 0.782 | $-8.74\times10^{-4}$ | $3.86\times10^{-7}$ | 0.997 | 13.7 |
| | 0.9 | 67.4 | 45.0 | 0.782 | $-8.74\times10^{-4}$ | $3.86\times10^{-7}$ | 1.008 | 9.1 |
| | 0.95 | 41.0 | 41.0 | 0.750 | $-6.52\times10^{-4}$ | $2.18\times10^{-7}$ | — | — |
| | 1 | 27.2 | 27.2 | 0.712 | $-5.58\times10^{-4}$ | $1.54\times10^{-7}$ | — | — |
| $FeSe_{1-x}Te_x$ (200h-annealed) | 0.6 | 26.5 | 0 | 0.782 | $-8.74\times10^{-4}$ | $3.86\times10^{-7}$ | 0.937 | 14.4 |
| | 0.7 | 35.7 | 0 | 0.782 | $-8.74\times10^{-4}$ | $3.86\times10^{-7}$ | 0.942 | 14.8 |
| | 0.8 | 57.2 | 33.0 | 0.782 | $-8.74\times10^{-4}$ | $3.86\times10^{-7}$ | 1.064 | 12.8 |
| | 0.9 | 66.6 | 43.0 | 0.782 | $-8.74\times10^{-4}$ | $3.86\times10^{-7}$ | 1.010 | 9.1 |
| | 0.95 | 33.7 | 33.7 | 0.656 | $-5.17\times10^{-4}$ | $1.46\times10^{-7}$ | — | — |
| | 1 | 25.6 | 25.6 | 0.643 | $-4.68\times10^{-4}$ | $1.11\times10^{-7}$ | — | — |



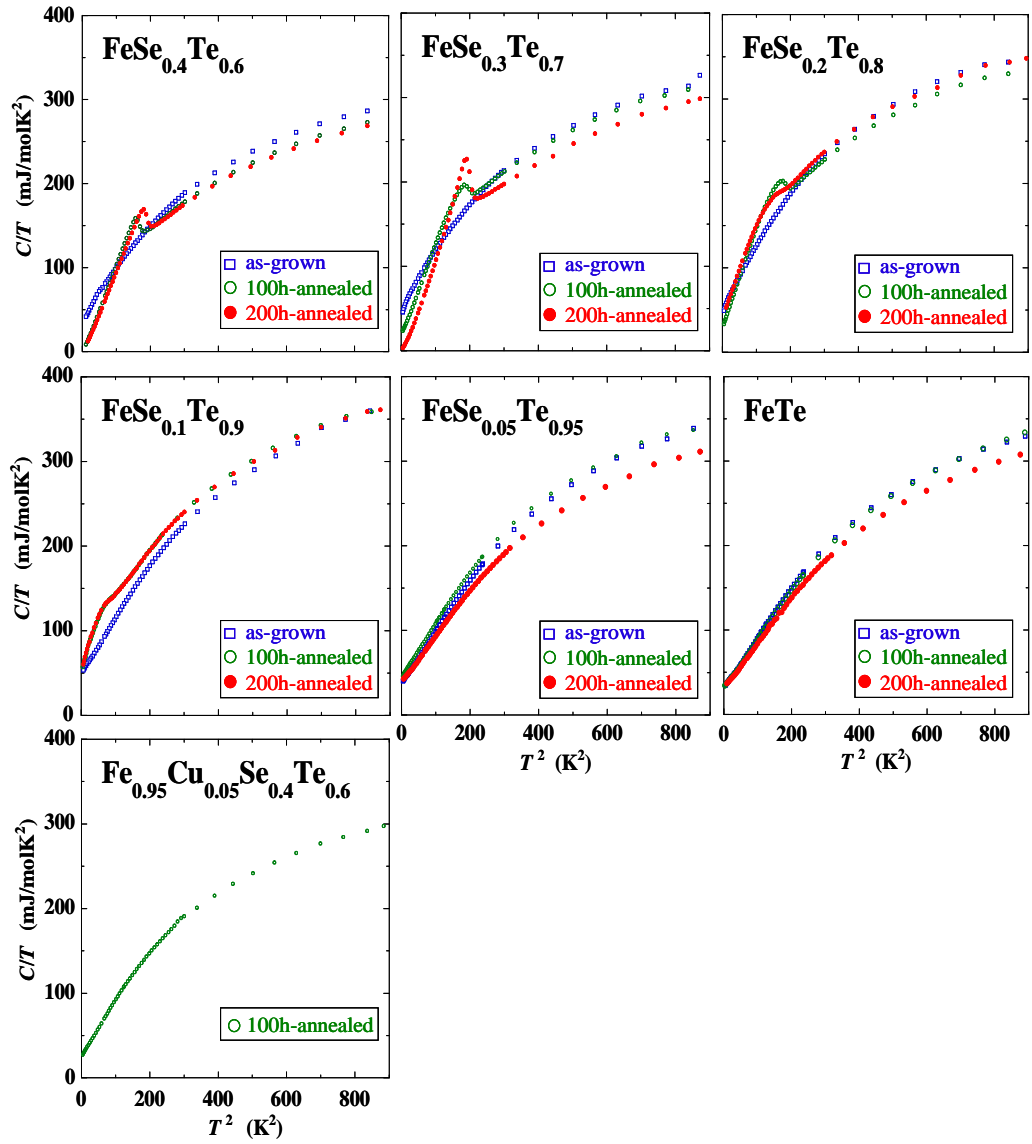

Fig. 1



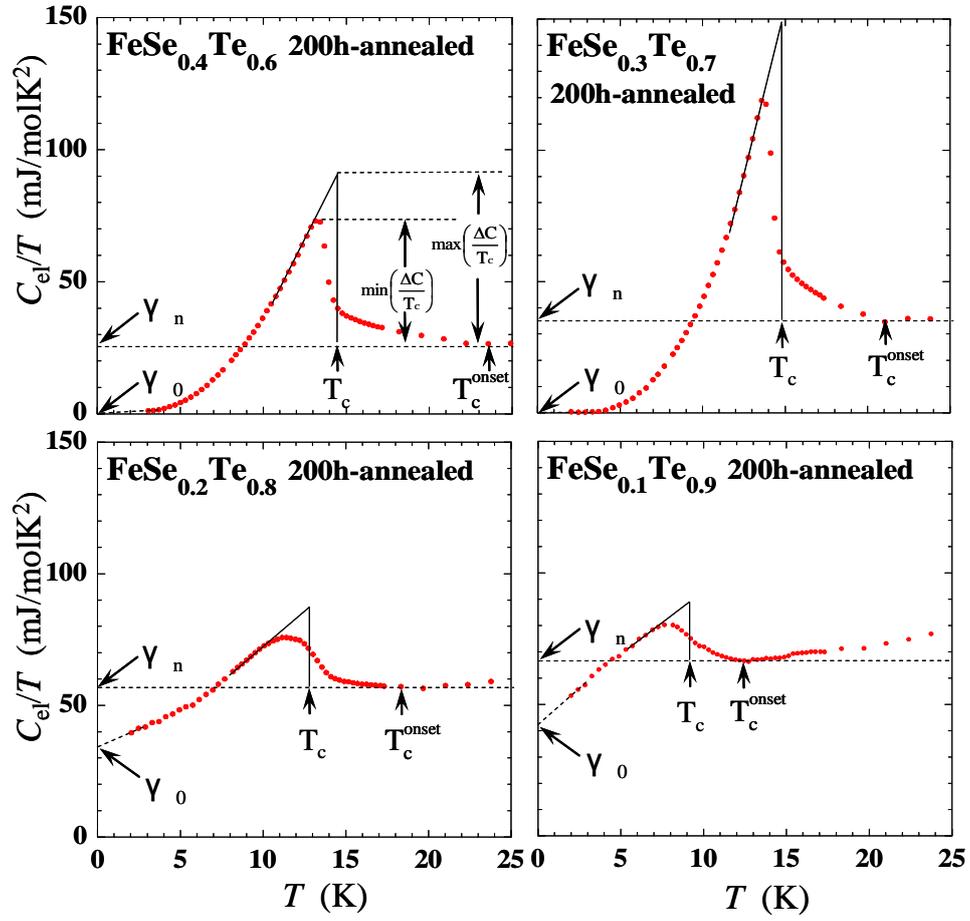



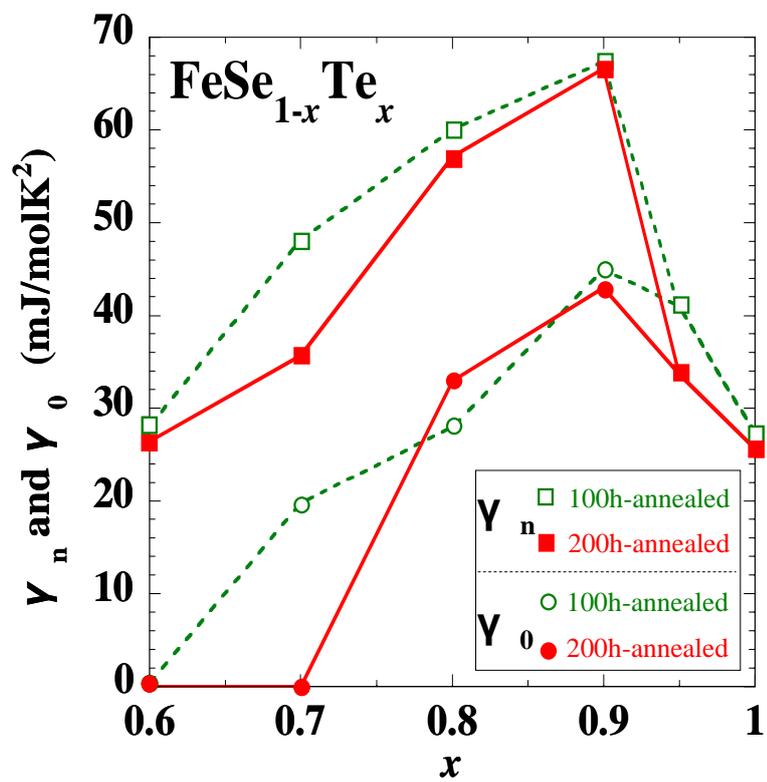

Fig. 3



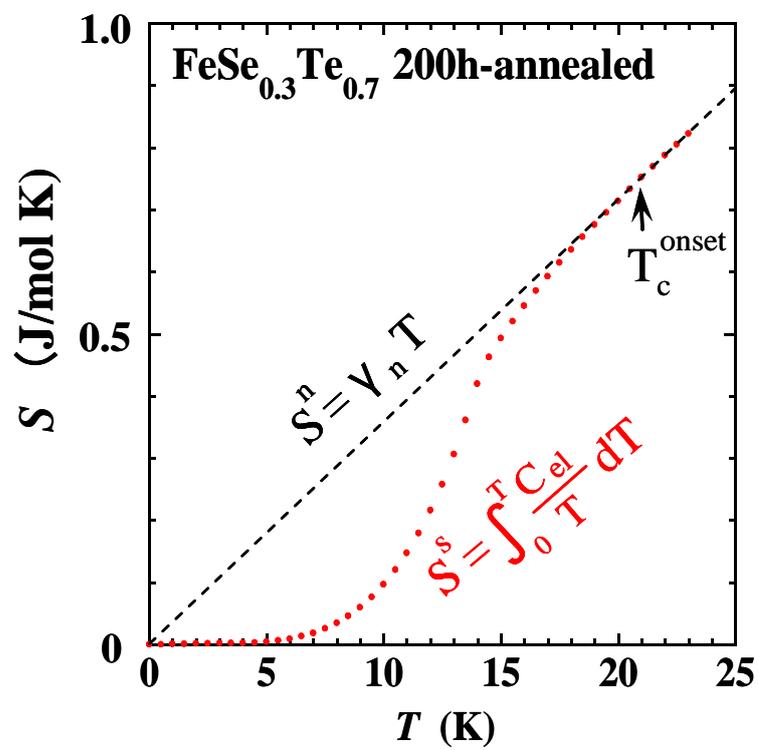

Fig. 4

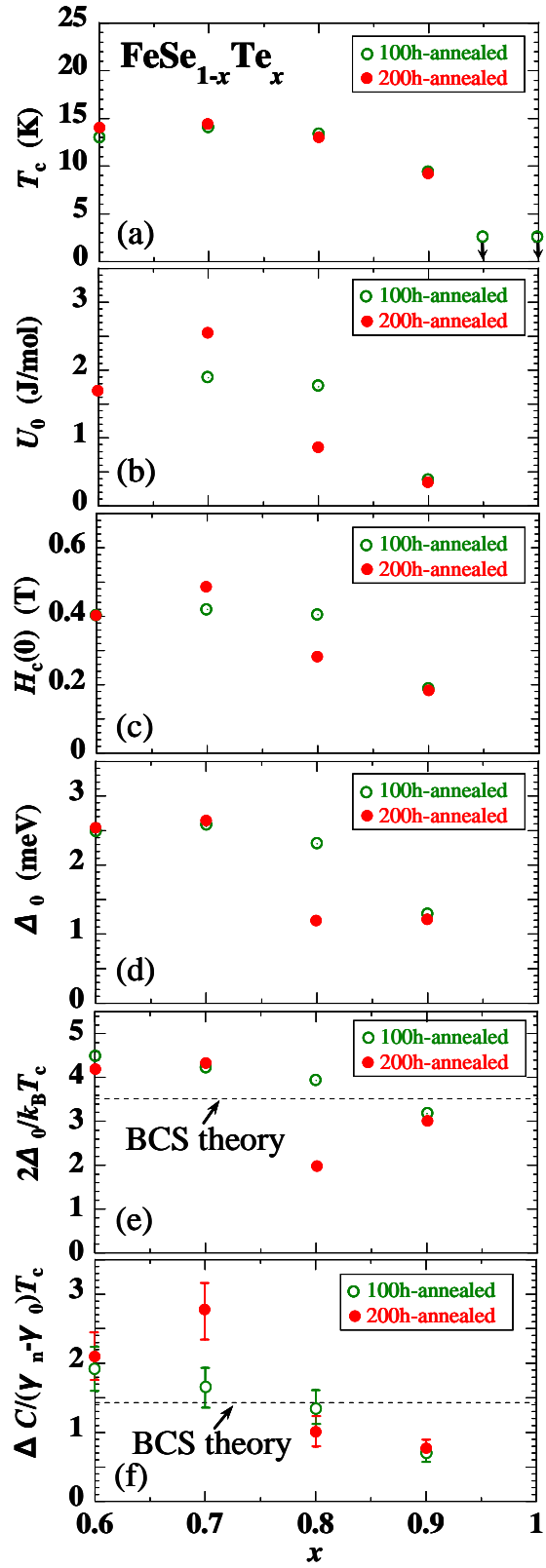

Fig. 5



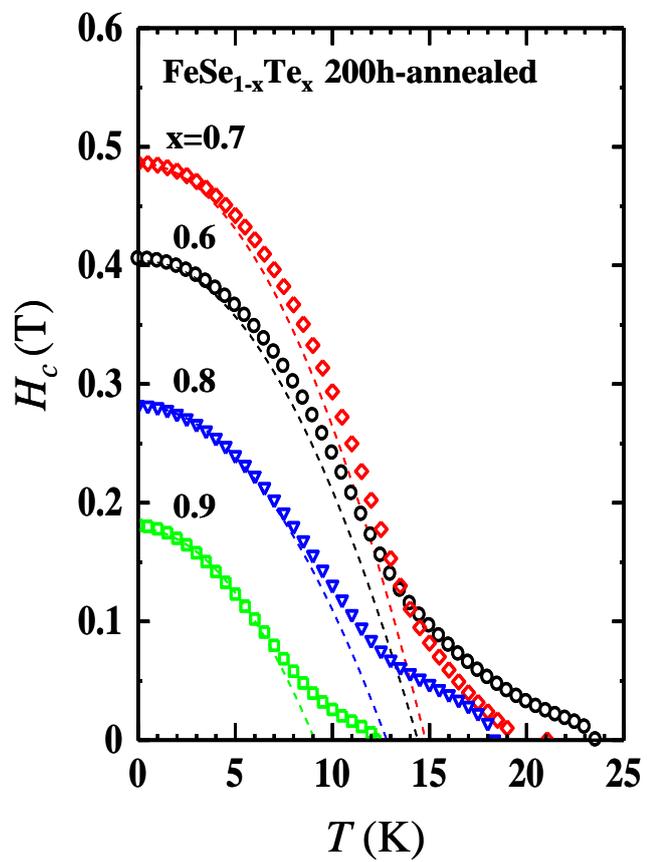

Fig. 6